\documentclass[fleqn,usenatbib]{mnras}

\usepackage{newtxtext,newtxmath}
\usepackage[T1]{fontenc}
\usepackage{ae,aecompl}

\usepackage{graphicx}	% Including figure files
\usepackage{amsmath}	% Advanced maths commands
\usepackage{siunitx}
\newcommand{\nustar}{\textit{NuSTAR}}
\newcommand{\suzaku}{{\it Suzaku}}
\newcommand{\swift}{{\it Swift}}
\newcommand{\xmm}{{\it XMM-Newton}}

\newcommand{\red}{\textcolor{black}}%{}%
\newcommand{\two}{\textcolor{black}}
\newcommand{\ergps}{erg~cm$^{-2}$~s$^{-1}$}
\newcommand{\src}{Mrk~1239}

\title[Mrk~1239 Spectral Analysis]{A Highly Accreting Low-Mass Black Hole Hidden in the Dust: \suzaku\ and \nustar\ observations of the NLS1 \src}

\author[J. Jiang et al.]{
Jiachen Jiang,$^{1}$\thanks{E-mail: jcjiang@mail.tsinghua.edu.cn} Mislav Balokovi\'{c}, $^{2,3}$ Murray Brightman,  $^{4}$ Honghui Liu,$^{5}$ \newauthor Fiona A. Harrison,$^{4}$ and George B. Lansbury$^{6}$
\\
$^{1}$Department of Astronomy, Tsinghua Univerisity, Shuangqing Road, Beijing 100084, China\\
$^{2}${Yale Center for Astronomy \& Astrophysics, 52 Hillhouse Avenue, New Haven, CT 06511, USA}\\ 
$^{3}${Department of Physics, Yale University, P.O. Box 2018120, New Haven, CT 06520, USA}\\
$^{4}$Cahill Center for Astronomy and Astrophysics, California Institute of Technology, Pasadena, CA 91125, USA\\
$^{5}$Department of Physics, Fudan University, 220 Handan Road, Shanghai 200433, China\\
$^{6}$European Southern Observatory, Karl-Schwarzschild Street 2, Garching bei M\"{u}nchen 85748, Germany\\
}

\date{Accepted XXX. Received YYY; in original form ZZZ}

\pubyear{2021}

\begin{document}
\label{firstpage}
\pagerange{\pageref{firstpage}--\pageref{lastpage}}
\maketitle

\begin{abstract}
We present {torus modelling} for the X-ray spectra of a nearby narrow-line Seyfert 1 galaxy \src\ ($z=0.0199$), based on archival \suzaku, \nustar\ and \swift\ observations. Our model suggests very soft intrinsic power-law continuum emission of $\Gamma\approx2.57$ in 2019 and $\Gamma\approx2.98$ in 2007. By applying a correction factor to the unabsorbed X-ray luminosity, we find that \src\ is accreting \red{near or around the Eddington limit}. Our best-fit spectral model also suggests a torus with a column density of $\log(N_{\rm H, ave}/$\,cm$^{-2})=25.0\pm0.2$ and a high covering factor of $0.90$ in \src, indicating that this source is most likely to be viewed almost \red{face-on} with $i\approx26^{\circ}$. Our line of sight might cross the edge of the torus with $N_{\rm H, los}=2-5\times10^{23}$\,cm$^{-2}$. The high Eddington ratio and the high line-of-sight column density makes \src\ one of the AGNs that are close to the limit where wind may form near the edge of the torus due to high radiation pressure. 
\end{abstract}


%%%%%%%%%%%%%%%%%%%%%%%%%%%%%%%%%%%%%%%%%%%%%%%%%%

%%%%%%%%%%%%%%%%% BODY OF PAPER %%%%%%%%%%%%%%%%%%

\section{Introduction} \label{intro}

\src\ is a narrow-line Seyfert 1 galaxy (NLS1) that shows broad Balmer components of {1000\,km\,s$^{-1}$ } and strong Fe [\textsc{II}] emission in the optical band \citep{osterbrock85, veron01}. \citet{ryan07} estimated the mass of the supermassive black hole (BH) in the center of \src\ to be $7.8\times10^{5}$\,$M_{\odot}$ {by using the size of the broad line region} based on the FWHM($\rm H{\beta}$)–$L_{\lambda 5100}$ relation \citep{kaspi05} and $1.3\times10^{6}$\,$M_{\odot}$ on the basis of the FWHM($\rm H{\beta}$)–$L_{\rm H{\beta}}$ relation \citep{green06}. By studying host bulge properties, \citet{graham07} obtained a similar BH mass of $5\times10^{5}-7\times10^{6}$\,$M_{\odot}$.

In addition to a low BH mass, \src\ also shows interesting properties across multiple wavelengths: in the radio band, relatively stronger radio emission than typical NLS1s along with evidence of a kiloparsec-scale radio jet was found \citep{doi15}, although \src\ is identified as a radio-quiet source \citep[e.g. 50\,mJy at 20\,cm,][]{ulvestad95}.  These features were previously seen only in radio-loud NLS1s \citep[e.g.][]{anton08, doi12}. In the near-infrared band, strong blackbody-like emission of $T\approx1200$\,K peaking at \SI{2.2}{\micro\metre} was found \citep{rodriguez06}. Such strong thermal emission might be related to the heated dusty torus {located between the narrow line region and the broad line region} of this system with a temperature close to the sublimation limit \citep{rodriguez06,riffel06}. 
%The other NLS1 Mrk~766 shows similar features peaking at \SI{2.25}{\micro\metre} but with much lower intensity {compared to \src} \citep{riffel06}. 
{In the optical band, polarisation studies by \citet{smith04} suggest that the optical band of \src\ is dominated by polar-scattered emission. Our line of sight (LOS) towards the nuclei should pass through the upper layer of the torus.}

In the X-ray band, \src\ shows particularly soft continuum emission. \citet{rush96} analysed the \textit{ROSAT} spectrum of \src\ between 0.1--2.4\,keV and found the soft X-ray spectrum was consistent with an absorbed power law with $\Gamma=2.94\pm0.04$. A later study by \citet{grupe04} suggested a similar conclusion of $\Gamma\approx3$ by analysing the \xmm\ observation of this source {in the 0.3--10\,keV band}. Such soft X-ray continuum emission indicates a high accretion rate in the central disc \citep[e.g.][]{brightman13}. An Eddington ratio of $\lambda_{\rm Edd}\approx2$ was obtained by modelling a multi-wavelength SED of \src\ and assuming $M_{\rm BH}=5\times10^{6}$\,$M_{\odot}$ \citep{grupe04}. 

\citet{grupe04} discovered an emission line feature at 0.9\,keV with an equivalent width of approximately $110$~eV in the same \xmm\ observation, which was interpreted as Ne~\textsc{ix} emission line in their work. {A super-solar Ne/O abundance might be required to explain this line feature. 

Broad band X-ray spectral analyses suggest that there are two light paths from \src: one of the light paths is absorbed direct emission while the other is less absorbed \citep{grupe04}, which is consistent with the discovery of wavelength-dependent polarisation degree in the optical emission of \src\ \citep{goodrich89}.} 

In this work, we present a broad-band spectral model for the \suzaku\ and \nustar\ observations of \src. In Section \ref{data}, we introduce our data reduction processes; in Section \ref{analysis}, we introduce a torus-based X-ray spectral model for \src; in Section \ref{discuss}, we discuss and conclude our results.

\section{Data Reduction} \label{data}

\subsection{\suzaku}

\textit{Suzaku} observed Mrk~1239 in 2007 for 63\,ks (obs ID: 702031010). We produce cleaned event files for all operating XIS detectors (0, 1, 3) using AEPIPELINE v1.1.0 and the latest CALDB as of 2019 November. Source extraction regions are chosen to be circles with radii of 120 arcsec and background are taken from nearby regions with the same shape. We use tasks XISRMFGEN and XISSIMARFGEN to create response files for each detector.  The spectra and response files of the front-illuminated instruments (XIS 0 and 3) spectra are combined by using the ADDSPEC tool. The combined spectrum is called FI spectrum hereafter. The spectrum of the back-illuminated instrument (XIS 1) is called BI spectrum hereafter. The spectra are grouped to have a minimum number of 20 counts per bin. During the spectral modelling, we ignore the energy band below 0.5\,keV and the 1.7--2.5\,keV band due to calibration uncertainty. 

\subsection{\nustar}

\src\ was observed by the \nustar\ satellite in 2019 for $\approx21$\,ks (observation ID 60360006002). The \nustar\ data are reduced using the standard pipeline NUPIPELINE v1.9.0 and instrumental responses from \nustar\ caldb V20200510. We extract the source spectra from circular regions with radii of 70\,arcsec, and the background spectra from nearby circular regions of 110\,arcsec on the same chip. The tool NUPRODUCTS is used for this purpose. The 3-40\,keV band is considered for both FPMA and FPMB spectra. {The energy band above 40\,keV is dominated by background.} The FPM spectra are grouped to have a minimum number of 20 counts per bin.

\subsection{\swift}

A \swift\ observation of \src\ with a length of 6\,ks that was taken simultaneously with our \nustar\ observation is also considered (observation ID: 00081986001). The calibration file version used for XRT data reduction is 20190412. The standard pipeline XRTPRODUCTS v0.4.2 is used for data processing. The source spectrum is extracted from a circular region with a radius of 40 arcsec and the background spectrum is extracted from a circular region with a radius of 100 arcsec nearby. The spectrum is binned to have a minimum count of 20 per bin. We consider the 0.5--6\,keV band of the XRT spectrum.

All the spectral analysis is processed by XSPEC v12.11.01 \citep{arnaud} in HEASOFT v6.27.2. $\chi^2$-statistics is considered in this work. The Galactic column density towards \src\ is fixed at the nominal value $4.16\times10^{20}$\,cm$^{-2}$ \citep{willingale13} if not specified. 

\begin{figure}
    \centering
    \includegraphics[width=\columnwidth]{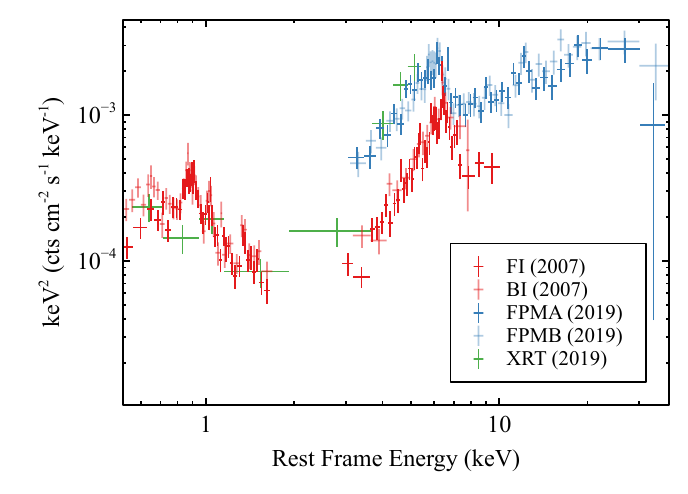}
    \caption{Unfolded \suzaku, \swift\ and \nustar\ spectra of \src. The response is unfolded using a featureless, constant model across the full band only for demonstration purposes. Blue: FPMA; faint blue: FPMB; red: FI; faint red: BI; green: XRT.}
    \label{pic_eeuf}
\end{figure}

\section{Spectral Analysis} \label{analysis}

Fig.\,\ref{pic_eeuf} shows the \suzaku, \swift\ and \nustar\ spectra of \src. A featureless, constant model is used to unfold the spectra only for demonstration purposes. The broad-band spectra of \src\ show similar obscuration as in other typical obscured Seyfert active galactic nuclei \citep[AGNs, e.g.][]{balokovic18}. 

In the rest of this section, we first present an overlook {of} the spectral properties of \src, and then we present a torus-based X-ray spectral model for the data.

\subsection{Iron Line and Compton Hump}

\begin{figure}
    \centering
    \includegraphics[width=\columnwidth]{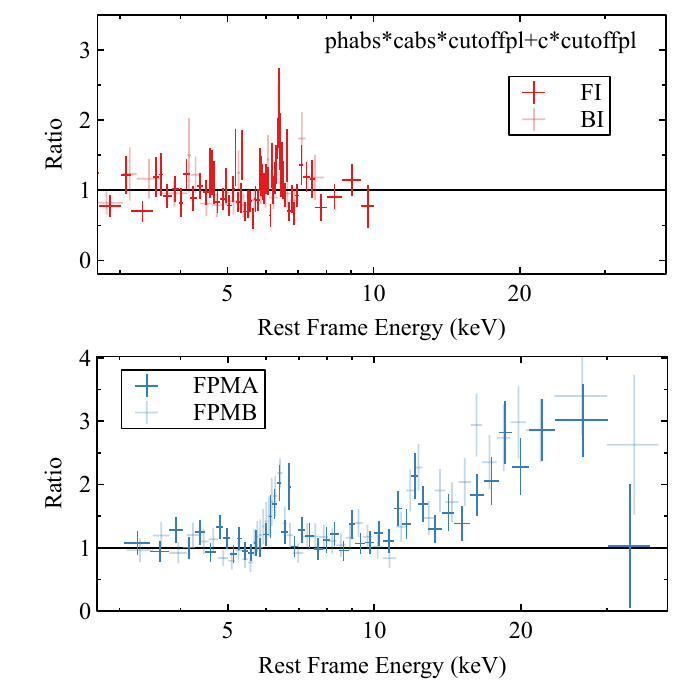}
    \caption{Data/model ratio plots using an absorbed power-law model plus a scattered power-law component. Both \suzaku\ and \nustar\ observations show a narrow emission line feature at the rest frame 6.4\,keV. FPM spectra also show {a} strong Compton hump feature above 10\,keV.}
    \label{pic_pl}
\end{figure}

Previous studies suggest that the emission of \src\ consists of two parts, one direct absorbed emission and one less absorbed scattered emission \citep{goodrich89,grupe04}. Therefore, we first model the \suzaku\ spectra of \src\ in the 2.5--10\,keV band with an absorbed power-law model plus a scattered power-law model. \red{The full model is \texttt{constant1 * tbnew * zmshift * ( vphabs * cabs * cutoffpl1 + constant2 * cutoffpl2 )}. The first \texttt{constant1} is used to account for calibration uncertainty of different instruments. The second \texttt{constant2} is the scaling factor ($f_{\rm S}$) of the scattered power-law component (\texttt{cutoffpl2}) relative to the direct obscured power-law component (\texttt{cutoffpl1}).}  The normalisation parameters and the photon index ($\Gamma$) of these two components are linked. The \texttt{zmshift} model is used to account for the source redshift ($z$=0.0199). The \texttt{tbnew} model accounts for Galactic absorption. The \texttt{vphabs} and \texttt{cabs} models are used to account for additional line-of-sight column density at the source's redshift.  

Such a model can mostly describe the continuum emission in the hard X-ray band very well. A line-of-sight column density of $5\times10^{23}$\,cm$^{-2}$ is required. See the top panel of Fig.\,\ref{pic_pl} for the corresponding data/model ratio plot. A narrow emission line feature was found peaking at the rest frame 6.4\,keV, which is the Fe K$\alpha$ emission from a cold emitter.

We apply the same model to the 3--10\,keV band spectra of the \nustar\ observation. The data/model ratio plots are shown in the lower panel of Fig.\,\ref{pic_pl}. \nustar\ spectra also show narrow Fe K$\alpha$ emission and a strong Compton hump above 10\,keV.

The existence of narrow Fe K$\alpha$ emission and Compton hump in the X-ray spectra suggest a cold, neutral emitter in this obscured system. Based on the evidence of an obscured continuum emission, we propose a torus interpretation \citep{antonucci93, urry95} for the X-ray reflector in \src.

\subsection{Torus Modelling}

\begin{figure}
    \centering
    \includegraphics[width=\columnwidth]{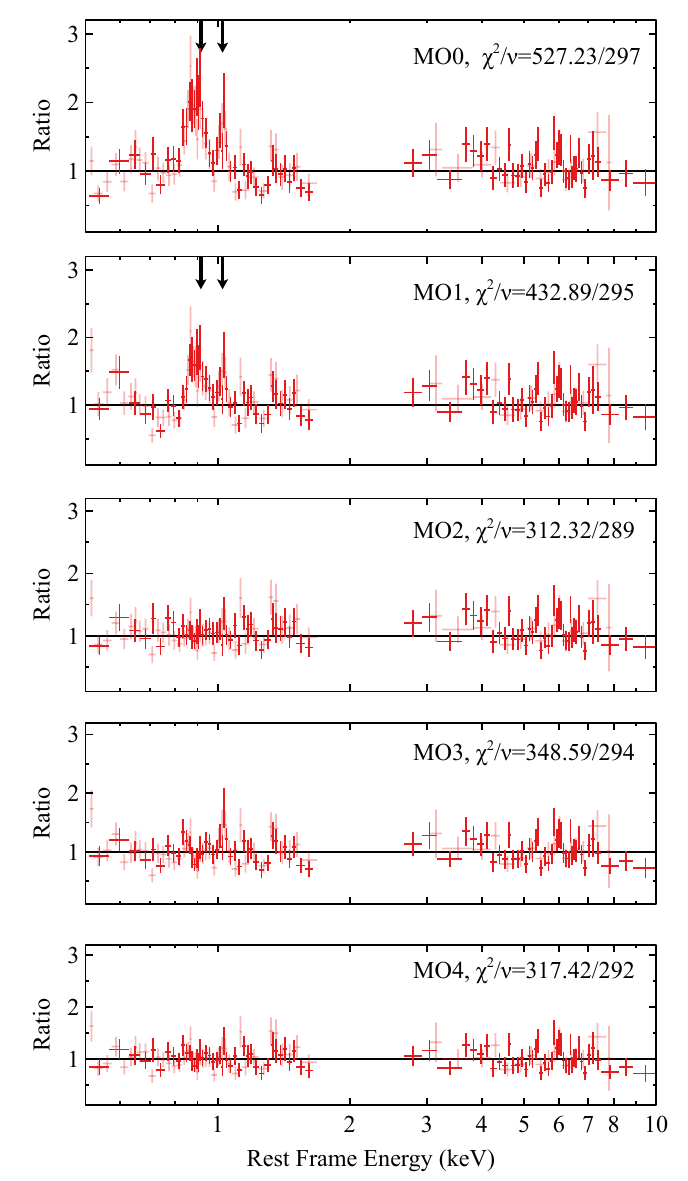}
    \caption{Data/model ratio plots of XIS spectra using model M0-M4. FI and BI spectra show significant evidence for an emission line feature at 0.9\,keV and tentative evidence for a second line at 1.0\,keV (marked by black arrows). They correspond to Ne~\textsc{ix} and Ne~\textsc{x} lines respectively.}
    \label{pic_gauss}
\end{figure}

{In this section, we model the full band spectra of \src\ using the \texttt{borus02} model introduced in \citet{balokovic18}.} 

\red{The \texttt{borus02} model represents reprocessed X-ray radiation from an approximately toroidal geometry originally proposed by \citet{brightman11}. It self-consistently accounts for the continuum and the fluorescent emission line components. The main parameters relevant for the spectral shape of the reprocessed component in the \texttt{borus02} model are the torus covering factor, its average column density, inclination and the relative abundance of iron. Despite the simplicity of the assumed geometry, these parameters form a complex and partially degenerate parameter space. The model is similar to, and broader than, the torus model of \citet{brightman11}, which has been shown to be incorrect \citep{liu15,balokovic18}. In this work we make use of the table model `borus02\_v200623sa.fits', which was calculated specifically to extend the photon index parameter space to accommodate sources with very steep intrinsic continua, such as \src.}

\red{In Appendix \ref{mytorus}, we present a spectral model for \src\ using the \texttt{mytorus} model, an alternative model for torus emission \citep{murphy09}. The spectra of \src\ require a very soft power-law continuum, which is beyond the allowed parameter range of the public version of \texttt{mytorus}. So, we only present the analysis using the \texttt{borus02} model in this section.}

\red{The full model is \texttt{constant1 * tbnew * zmshift * ( borus02 + vphabs * cabs * cutoffpl1 + constant2 * cutoffpl2 )} (MO0) in the XSPEC format.} 

During our spectral fitting process, the photon index ($\Gamma$), the high-energy cut-off ($E_{\rm cut}$) and the normalisation parameters of the \texttt{borus02} model are linked to the corresponding parameters of \texttt{cutoffpl1} and \texttt{cutoffpl2}. {Previous optical polarisation studies suggest our LOS might cross the upper layer of the torus in \src} \citep[e.g.][]{smith04}. We thus tie the half-opening angle and the inclination angle of the torus in our analysis. Similar approach was taken in the X-ray data analysis of other obscured AGNs \citep[e.g.][]{kamraj19}. Later in Section \ref{sec_i}, we will discuss the values of these two parameters in detail. {The abundance parameters of \texttt{vphabs} model is linked to the iron abundance of \texttt{borus02} by using the solar abundances calculated from \citet{anders89}.} 

\begin{figure*}
    \centering
    \includegraphics[width=17cm]{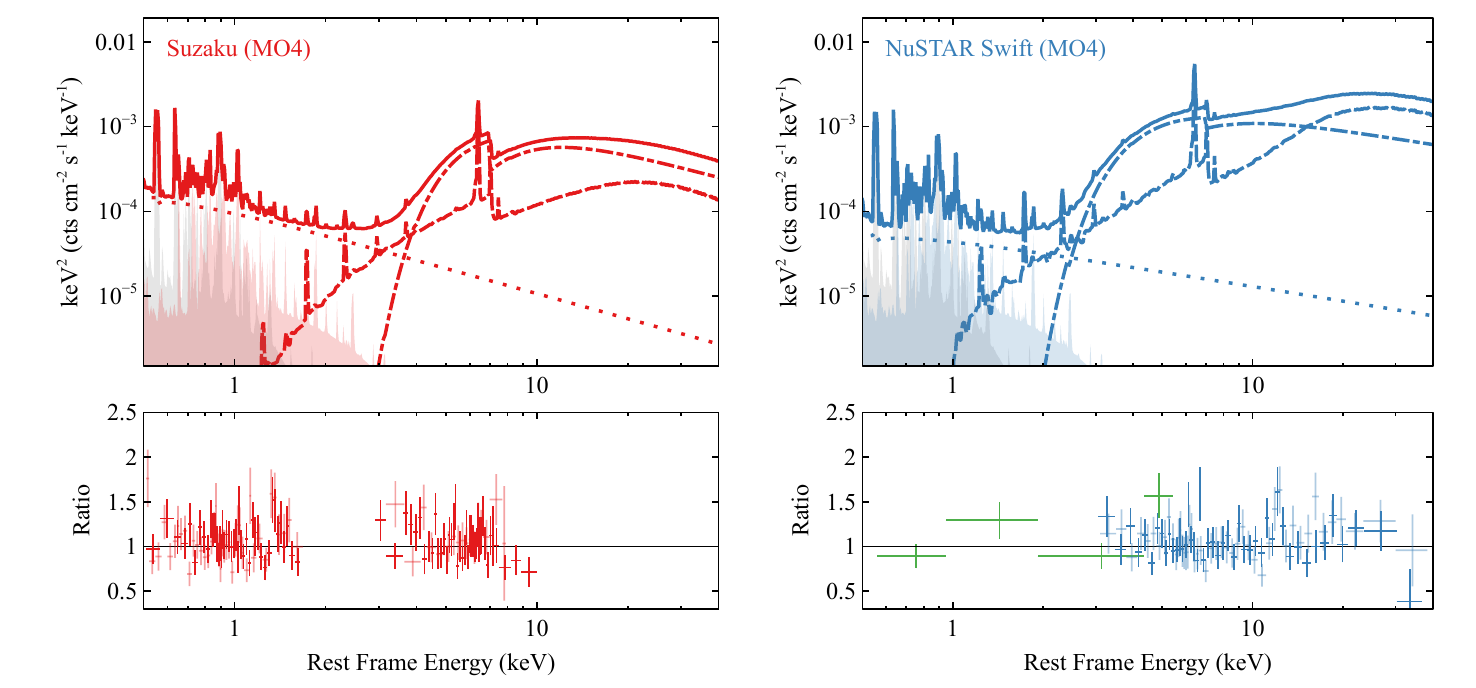}
    \caption{Top: best-fit models for the \suzaku\ (left), \nustar\ and \swift\ (right) spectra of \src\ using MO4. Thick lines: total model; shaded regions: two diffuse hot plasma emission components; dashed lines: torus model; dash-dotted lines: absorbed power-law continuum emission; dotted lines: scattered power-law component.s Bottom: corresponding data/model ratio plots.}
    \label{pic_final}
\end{figure*}

MO0 is able to provide a reasonable fit for the FI and BI spectra above 2.5\,keV. But the fit is less satisfying in the soft X-ray band with $\chi^{2}/\nu=527.23/297$. See the first panel of Fig.\,\ref{pic_gauss} for the corresponding data/model ratio plot. %Only an absorbed scattered power-law component is unable to fit the soft X-ray emission.

\subsection{Diffuse Plasma Emission in the Soft X-ray Band} \label{sec_soft}

\subsubsection{Ne~\textsc{x} and Ne~\textsc{ix} Emissions}

{In order to fit the soft X-ray emission better, we include a diffuse hot plasma model \texttt{vmekal} \citep{liedahl95} by following the same approach for typical obscured Seyfert AGNs \citep[e.g.][]{itoh08,hernandez17}. In the beginning, we fix the abundances of \texttt{vmekal} at solar values \citep{anders89}.} The full model is \texttt{constant1 * tbnew * zmshift * ( borus02 + vmekal + vphabs * cabs * cutoffpl1 + constant2 * cutoffpl2 )} (MO1). MO1 significantly improves our fit of the \suzaku\ spectra of \src\ in the soft X-ray band with $\chi^{2}=432.89/298$. See the second panel of Fig.\,\ref{pic_gauss}. 

However, two emission line features are still seen around 1\,keV. The first emission line feature is at the rest frame 0.9\,keV, which is similar to the Ne~\textsc{ix} line identified in the \xmm\ observation \citep{grupe04}. A second emission line is found around 1\,keV. We first model these emission lines by using two simple Gaussian line models. The full model is \texttt{constant1 * tbnew * zmshift * ( gauss1 + gauss2 + borus02 + vmekal + vphabs * cabs * cutoffpl1 + constant2 * cutoffpl2 )} (MO2).

By fitting the first line around 0.9\,keV with \texttt{gauss1}, the fit is improved by $\Delta\chi^{2}=104$ and 3 more free parameters. The best-fit parameters are shown in Table\,\ref{tab_fit}. This line is at $0.884\pm0.007$\,keV, and lies at the energy of Ne~\textsc{ix} emission. The equivalent width of the line is $117^{+23}_{-12}$\,eV. The best-fit values are consistent with previous \xmm\ measurements \citep{grupe04}. 

By fitting the second line around 1\,keV with \texttt{gauss2}, we are able to improve the fit by $\Delta\chi^{2}=17$ and 3 more free parameters. The line is at $1.024^{+0.010}_{-0.012}$\,keV {in the rest frame} and the equivalent width is $52^{+14}_{-23}$\,eV. We only obtain an upper limit of its line width ($\sigma<0.02$\,keV). This narrower emission line at 1.03\,keV can be interpreted as Ne~\textsc{x} emission line. A quick F-test based on the $\chi^{2}$ improvement suggests an F statistic value of 7.7, which is much less significant than the Ne~\textsc{ix} line. We conclude that we find tentative evidence for Ne~\textsc{x} emission line in addition to strong evidence of Ne~\textsc{ix} line as in previous analyses.

\subsubsection{Hot Diffuse Plasma with Super-solar Ne Abundances}

\red{The two narrow emission lines lie well with the energy of Ne~\textsc{x} and Ne~\textsc{ix} lines. As \citet{grupe04} argued, super Ne abundances relative to oxygen might be needed to explain these lines. Instead of modelling the lines with simple \texttt{gauss} models, we propose a physical model with a super-solar Ne abundance for the data. To do so, the Ne abundance parameter ($Z_{\rm Ne}$) of the \texttt{vmekal} component is allowed to be free during spectral fitting (MO3). Other abundances are fixed at solar \citep{anders89}.}  

\red{MO3 is able to provide a better fit than MO1 with $\Delta\chi^{2}=84.3$ and one more parameter. Corresponding data/model ratio plots are shown in the fourth panel of Fig.\,\ref{pic_gauss}. Best-fit parameters are shown in Table \ref{tab_fit}. MO3 offers a reasonable fit to the emission line at 0.88\,keV with a super-solar Ne abundance, although some residuals still remain around 0.86~keV. The second emission line at 1.02~keV cannot be modelled by MO3. Therefore, we add a second \texttt{vmekal} component (MO4). The Ne abundances of these two \texttt{vmekal} components are linked. MO4 provides a very good fit of the data with $\chi^{2}/\nu=317.42/292$. The best-fit parameters are shown in Table\,\ref{tab_fit}, and corresponding data/model ratio plots are shown in the last panel of Fig.\,\ref{pic_gauss}.}

\begin{figure}
    \centering
    \includegraphics{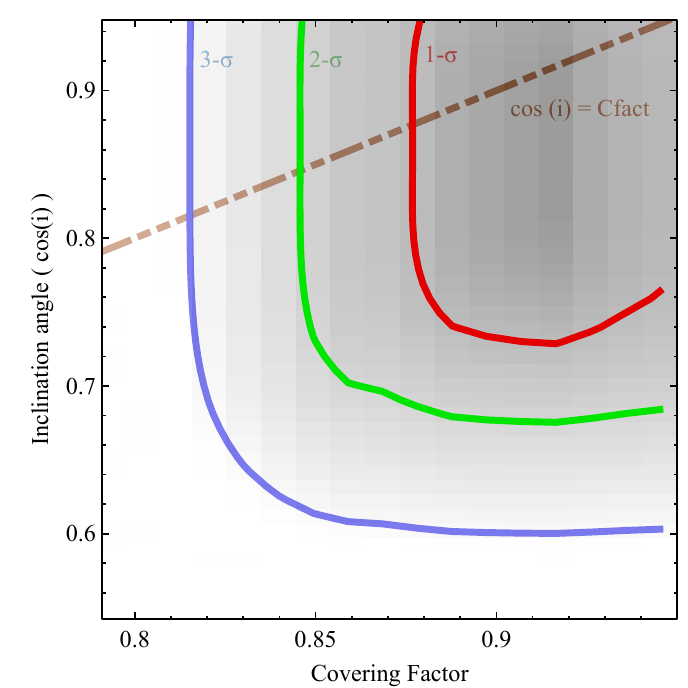}
    \caption{$\chi^{2}$ distribution on the Cfact and $i$ parameter plane. The red, green and blue solid lines show the 1$\sigma$, 2$\sigma$ and 3$\sigma$ uncertainty ranges.The dash-dotted line shows the Cfact=$\cos(i)$ line.}
    \label{pic_i}
\end{figure}

\red{In comparison with MO2 where \texttt{gauss} models are used, the fit using MO4 has a slightly higher $\chi^{2}$ ($\Delta\chi^{2}=5.1$) but 3 fewer parameters. In the end, we decide to choose MO4 as our best-fit model instead of MO2. Because MO4 provides a more physical interpretation to the narrow emission lines. }

\red{In conclusion, two diffuse plasma components are required to fit the soft X-ray spectrum of \src: one has a temperature of $kT\approx0.22$\,keV and the other one has a temperature of $kT\approx0.64$\,keV. A super-solar Ne abundance of $Z_{\rm Ne}=4.9^{+1.2}_{-1.3}Z_{\odot}$ is needed. \citet{buhariwalla20} suggested that these hot plasma components may be associated with the star-burst region in the host galaxy of \src. \two{This component has been often found in the soft X-ray emission of many obscured sources \citep[e.g.][]{franceschini03}. When the AGN emission is high and not heavily obscured, this component still exists, but overwhelmed by the central AGN emission. }}

\subsection{Multi-Epoch Spectral Analysis}

In this section, we apply MO4 to all the spectra of \src\ simultaneously. As shown in Fig.\,\ref{pic_eeuf}, the soft X-ray band of \src, which is dominated by distant diffuse plasma emission, shows a consistent flux level in two epochs. The signal-to-noise (S/N) of \swift\ XRT data does not allow us to investigate the soft X-ray emission during the second epoch in detail. Therefore, we link the parameters of diffuse plasma emission components for two epochs. Other parameters, such as the torus inclination angle, the average column density and the iron abundance of the torus, are expected to be consistent on observable timescales. They are thus linked during our spectral fitting too.

\red{MO4 provides a good fit to the data of both epochs. The best-fit parameters are shown in Table\,\ref{tab_final}, and the best-fit data/model ratio plots are shown in Fig.\,\ref{pic_final}. }

\subsubsection{The Inclination Angle and the Opening Angle of the Torus} \label{sec_i}

\red{The best-fit average column density of the torus $\log(N_{\rm H, ave}$ is approximately $10^{25}{\rm cm^{-2}}$. The line-of-sight column density $N_{\rm H, los}$ is 2--5$\times10^{23}$\,cm$^{-2}$, which is 2 orders of magnitudes lower than $N_{\rm H, ave}$. In MO4, the inclination angle ($i$) and the half-opening angle\footnote{The covering factor of a toroid-shaped torus is approximately the cosine of its half-opening angle.} of the torus (Cfact) are linked in our analysis as in \citet{kamraj19} -- our LOS might cross the edge of the torus where the column density is only 2--5\% of the average column density of the whole torus.}

\red{In this part of the section, we investigate how valid the Cfact$\approx\cos(i)$ relation is by allowing these two parameters to be free in our analysis. All the spectra are used for this test. Fig.\,\ref{pic_i} shows the $\chi^{2}$ distribution on the Cfact and $i$ parameter plane. The XSPEC tool STEPPAR is used to do so. Only lower limit has been obtained: a 3$\sigma$ uncertainty range of Cfact is >0.815 and that of $\cos(i)$ is >0.6. They suggest a half-opening angle smaller than $35^{\circ}$ and an inclination angle smaller than $53^{\circ}$. The model is more sensitive to Cfact/the half-opening angle of the torus, because this parameter directly modifies the continuum emission. The constraint of $i$ is relatively weaker as the model is less sensitive to this parameter.} 

\red{We show the Cfact=$\cos(i)$ reference line in Fig.\,\ref{pic_i}. Our assumption of Cfact=$\cos(i)$ lines well within the 1$\sigma$ uncertainty range, indicating that these two parameters approach a similar value in our analysis--our LOS might cross the edge of a torus which has a small half-opening angle. Similar conclusions were found in the optical polarisation studies: \src\ turns out to be a polar-scattered NLS1 in which our LOS passes the upper layer of its torus \citep{smith04}.}

\red{By linking these two parameters, we obtain a better constraint of both parameters Cfact=$\cos(i)=0.90^{+0.04}_{-0.03}$ (90\% confidence range) as shown in Table\,\ref{tab_final}. This best-fit value corresponds to a small half-opening and inclination angle of approximately 26$^{\circ}$.}

\red{A covering factor that is as high as $\approx90\%$ in \src\ has been also seen in \nustar\ observations of other obscured Sy1 galaxies but with lower accretion rates than in \src\ \citep[e.g.][]{balokovic18,kamraj19}.}

\subsubsection{Ultra-soft X-ray Continuum Emission}

The photon index of the power-law continuum emission is $2.98\pm0.02$ for the \suzaku\ epoch and $2.57^{+0.03}_{-0.02}$ for the \nustar\ and \swift\ epoch. Despite of the obscured nature, the very soft X-ray continuum emission from the hot corona of \src\ is very similar to that of other NLS1s \citep[e.g.][]{gallo18}. 

In particular, X-ray studies of a sample of unobscured, extreme ultra-soft NLS1s suggest similar continuum emission with $\Gamma>2.5$ \citep{jiang20}. Detailed modelling of their multi-wavelength SEDs suggest an Eddington ratio that is around or a few times higher than the Eddington limit \citep[e.g.][]{jin09, jiang20}. Strong soft excess emission is shown in their data, and can be interpreted as part of reflection from a highly ionised inner disc region as well broad Fe K emissions \citep{jiang20}. Unfortunately, due to the obscuration along our LOS towards \src, we are unable to constrain the soft excess emission from the center of the AGN in \src. 

In Section\,\ref{relxill}, instead of investigating in the soft X-ray band, we discuss possible contribution of a disc reflection component in the hard X-ray band, particularly in the observed back-scattering Compton hump.

\begin{table*}
    \centering
    \begin{tabular}{ccccccc}
    \hline\hline
    Model & Parameter & Unit & MO2 & MO3 & MO4 \\
    \hline
    \texttt{tbnew} & $N_{\rm H}$ & $10^{20}$\,cm$^{-2}$ & $3.9^{+0.4}_{-0.4}$ & $4.1^{+0.3}_{-0.4}$ &  $3.8\pm0.6$ \\
    \hline
    \texttt{borus02}  & $\log(N_{\rm H, ave})$ & cm$^{-2}$ & $25.0^{+0.4}_{-0.3}$ & $25.3^{+0.2}_{-0.4}$ & $25.0^{+0.4}_{-0.3}$ \\
    & Cfact & - & $0.91^{+0.03}_{-0.05}$ & $0.92^{+0.03}_{-0.04}$ & $0.92^{+0.02}_{-0.04}$ \\
    & $Z_{\rm Fe}$ & $Z_{\odot}$ & $1.4^{+0.3}_{-0.2}$ & $1.5\pm0.4$ & $1.6^{+0.4}_{-0.3}$ \\
    \hline
    \texttt{vphabs} & $N_{\rm H, los}$ & $10^{22}$\,cm$^{-2}$ & $55^{+6}_{-7}$ & $47^{+6}_{-4}$ & $48^{+4}_{-3}$ \\
    \hline
    \texttt{cutoffpl1,2} & $\Gamma$ & - & $2.98\pm0.02$ & $2.98\pm0.02$ & $2.98^{+0.02}_{-0.03}$ \\
    & $E_{\rm cut}$ & keV & 500 & 500 & 500 \\
    & norm & $10^{-2}$ & $2.2^{+0.4}_{-0.2}$ & $1.9^{+0.4}_{-0.5}$ & $1.6^{+0.4}_{-0.5}$ \\
    \hline
    \texttt{constant2} & $f_{\rm S}$ & $10^{-3}$ & $4.6\pm0.3$ & $10.7\pm0.3$ & $8.9\pm0.4$ \\
    \hline
    \texttt{vmekal1} & kT & keV & $0.21^{+0.02}_{-0.03}$ & $0.23^{+0.02}_{-0.03}$ & $0.22^{+0.02}_{-0.03}$\\
    & $Z_{\rm Ne}$ & $Z_{\odot}$ & 1 & $4.7^{+2.3}_{-1.0}$ & $4.9^{+1.2}_{-1.3}$ \\
    & norm & $10^{-4}$ & $1.16^{+0.07}_{-0.10}$ & $1.05^{+0.02}_{-0.04}$ & $1.07\pm0.03$ \\ 
    \hline
    \texttt{vmekal2} & kT & keV & - & - & $0.64^{+0.02}_{-0.03}$ \\
    & norm & $10^{-5}$ & - & - & $2.2\pm0.2$\\
    \hline
        \texttt{gauss1} & Eline & keV & $0.884\pm0.007$ & - & -\\
    & $\sigma$ & kev & $0.039^{+0.007}_{-0.010}$ & - & -\\
    & EW & eV & $117^{+23}_{-12}$ & - & -\\
    \texttt{gauss2} & Eline & keV & $1.024^{+0.010}_{-0.012}$ & - & -\\
    & $\sigma$ & kev & $<0.02$ & - & -\\
    & EW & eV & $56^{+14}_{-23}$ & - & -\\
    \hline
    & $\chi^{2}/\nu$ & & 312.32/289 & 348.59/294 & \two{317.42/292} \\
    \hline\hline
    \end{tabular}
    \caption{Best-fit model parameters for the \suzaku\ observation of \src. MO2 uses Gaussian line models to fit the narrow emission lines around 1\,keV in the data; MO3 considers one \texttt{vmekal} component with super-solar Ne abundances; MO4 considers two \texttt{vmekal} components with the same super-solar Ne abundances. The other abundances of the \texttt{vmekal} models are assumed to be solar \citep{anders89}.}
    \label{tab_fit}
\end{table*}

\begin{table}
    \centering
    \begin{tabular}{ccccccc}
    \hline\hline
    Model & Parameter & Unit & Su & Nu\&Sw \\
    \hline
    \texttt{tbnew} & $N_{\rm H}$ & $10^{20}$\,cm$^{-2}$ & $3.8\pm0.4$ & $l$  \\
    \hline
    \texttt{borus02}  & $\log(N_{\rm H, ave})$ & cm$^{-2}$ & $25.0\pm0.2$ & $l$ \\
    & Cfact & - & $0.90^{+0.04}_{-0.03}$ & $l$\\
    & $Z_{\rm Fe}$ & $Z_{\odot}$ & $1.3^{+0.4}_{-0.3}$ &  $l$\\
    \hline
    \texttt{vphabs} & $N_{\rm H, los}$ & $10^{22}$\,cm$^{-2}$ & $52^{+7}_{-6}$ & $20^{+4}_{-3}$ \\
    \hline
    \texttt{cutoffpl1,2} & $\Gamma$ & - & $2.98\pm0.02$ & $2.57^{+0.03}_{-0.02}$ \\
    & $E_{\rm cut}$ & keV & >480 & $l$\\
    & norm & $10^{-2}$ & $1.4\pm0.4$ & $0.6\pm0.3$ \\
    \hline
    \texttt{constant2} & $f_{\rm S}$ & $10^{-3}$ & $8.3\pm0.2$ & $7.4\pm0.3$ \\
    \hline
    \texttt{vmekal1} & kT & keV & $0.21\pm0.02$ & $l$\\
    & $Z_{\rm Ne}$ & $Z_{\odot}$ & $4.9^{+1.3}_{-1.0}$ & $l$ \\
    & norm & $10^{-4}$ & $1.10\pm0.04$ & $l$ \\ 
    \hline
    \texttt{vmekal2} & kT & keV & $0.63\pm0.02$ & $l$\\
    & norm & $10^{-5}$ & $3.0\pm0.2$ & $l$ \\ 
    \hline
    & $\chi^{2}/\nu$ & & 519.97/468 \\
    \hline\hline
    \end{tabular}
    \caption{Best-fit model parameters for two epochs. $l$ means this parameter is linked during our spectral fitting.}
    \label{tab_final}
\end{table}

\section{Discussion} \label{discuss}

\subsection{The Accretion Rate of \src}

Our best-fit model suggests a very soft intrinsic power-law emission of $\Gamma=2.6-3$. The softness of the continuum emission suggests a high mass accretion rate in the disc \citep[e.g.][{and references therein}]{brightman13}.

\red{We estimate the Eddington ratio of \src\ by using its 2--10\,keV luminosity. The absorption-corrected flux of \src\ is $1.0\times10^{-11}$\,\ergps\ in 2007 and $8.0\times10^{-12}$\,\ergps\ in 2019 calculated by our best-fit model. Assuming a BH mass of $1\times10^{6}$\,$M_{\odot}$, they correspond to $L_{\rm X}=7.1-8.8\times10^{42}$\,erg\,s$^{-1}=0.06-0.07$\,$L_{\rm Edd}$. When considering a typical correction factor of $\kappa=10-20$ for this luminosity \citep{vasudevan07}, we estimate \src\ has a bolometric luminosity of 0.6-1.4 of the Eddington limit assuming a BH mass of $10^{6}$\,$M_{\odot}$. }

\red{Previous independent measurements of the mass of the SMBH in \src\ all agree with a relatively low value of $\approx1\times10^{6}$\,$M_{\odot}$ (see Section\,\ref{intro}). We consider the largest measurement uncertainty in the literature: \citet{graham07} estimated $M_{\rm BH}=5\times10^{5}-7\times10^{6}$\,$M_{\odot}$ by using the host bulge properties in \src. After taking into account the uncertainty of BH mass measurements, our estimation of $\lambda_{\rm Edd}$ is 0.1--2.8 for \src.}   

\red{Similar conclusions were found in \citet{grupe04} where a multi-wavelength SED was used for the calculation of the Eddington ratio: $\lambda_{\rm Edd}\approx2$ assuming $M_{\rm BH}=5\times10^{6}$\,$M_{\odot}$. \citet{buhariwalla20} found an Eddington ratio of $\lambda_{\rm Edd}=1-1.5$. \citet{yao18} applied a correction factor to the $L_{\lambda\rm 5100}$ of \src\ and obtained $\lambda_{\rm Edd}=1.12$. The uncertainties of their estimations were not mentioned in their work. But they are all consistent with our measurement.}

\red{In conclusion, \src\ is one of the most extreme AGNs that are accreting near or around the Eddington limit.} 

\subsection{Stability of the Dusty Torus in \src}

The high accretion rate in the central engine of \src\ may explain the high temperature blackbody emission in the near-infrared band \citep{rodriguez06}. The dusty torus is heated by the radiation from the inner accretion region close to the sublimation limit. In this section, we discuss the stability of such a heavy torus and the radiation pressure from the luminous nucleus onto the torus. 

\begin{figure}
    \centering
    \includegraphics[width=\columnwidth]{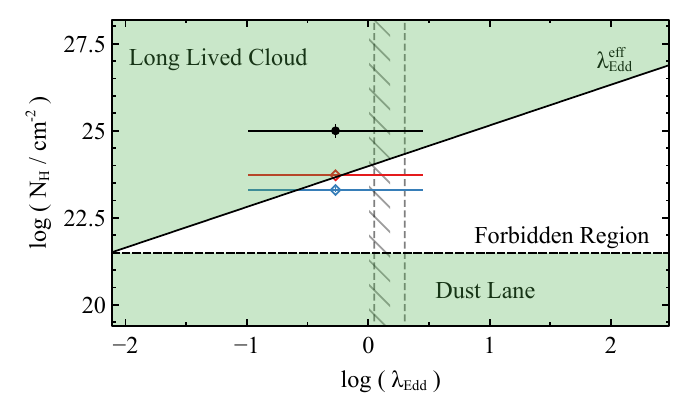}
    \caption{Eddington ratio $\lambda_{\rm Edd}$ vs. column density diagram. The solid black line shows the effective Eddington limit for different values of column density. The dashed line shows the typical column density of the dust lanes in host galaxies. Obscuring dusty materials will be blown away due to radiation pressure when luminosity exceeds the effective Eddington limit (white forbidden region in the figure). Most \swift\ BAT AGN lie within the green shaded region \citep{fabian09, ricci17}. Some exceptional dusty quasars at higher redshifts, e.g. $z>0.2$, are found to be located in the forbidden region \citep{lansbury20}. {The black circle shows the column density of the torus in \src. The red and blue diamonds shows the line-of-sight column density respectively in 2007 and 2019. The luminosity of the nuclei in \src\ may exceed the effective Eddington limit in the upper layer of the torus. Wind might be forming on the edge due to high radiation pressure. \red{The left and right grey dashed lines show the values $\lambda_{\rm Edd}$ given in \citet{yao18} and \citet{grupe04} respectively. The shaded region shows the range of $\lambda_{\rm Edd}$ estimated by \citet{buhariwalla20}}.}}
    \label{pic_nh}
\end{figure}

\citet{laor93,scoville95,murray05} calculate the effective Eddington ratio for cool dusty gas, which is found to be much lower than the Eddington ratio for ionised dust-free gas. As an example, the black solid line in Fig.\,\ref{pic_nh} shows the effective Eddington limit for different column densities. The line is adopted from \citet{fabian09}. The radiation pressure will play an important role for systems that correspond to the right of the solid line. The obscuring dusty materials will be blown away due to high radiation pressure when luminosity exceeds the effective Eddington limit. The dashed line shows the typical column density of galactic dust lane that may make contribution to the X-ray spectrum. 

Large X-ray surveys, e.g. the \swift\ BAT catalog, indeed find that most of the known AGNs tend to avoid the forbidden region of this diagram \citep{fabian09, ricci17} and lie within the green shaded region in Fig.\,\ref{pic_nh}. Some exceptional cases are, however, found in very luminous dusty quasars at higher redshifts \citep[e.g. $z>0.2$,][]{lansbury20}.

We show \src\ in the $N_{\rm H}-\lambda_{\rm Edd}$ diagram in Fig.\,\ref{pic_nh}, and it is located on the edge of the `long-lived cloud' region. The line-of-sight column density of \src\ is estimated to be $2-5\times10^{23}$\,cm$^{-2}$. The luminosity of \src\ can exceed the effective Eddington limit at this low value of column density. Therefore, radiation pressure-driven wind might be forming near the edge of the torus along our LOS. 

\red{Last but not least, it is important to note that the H$\alpha$ emission of \src\ shows an asymmetric profile with a minimum polarisation degree in the blue wing and a maximum degree in the red wing, which might be related to a radial outflow \citep{smith04}.}

Future high-S/N, TES-based (Transition Edge Sensor) observations, such as from \textit{XRISM} with a spectral resolution of 2.5\,eV, might enable us to look for any evidence of wind absorption features in the X-ray spectra, such as blueshifted Fe K edges.

\subsection{The Variability of the Line-of-Sight Column Density of \src}

\two{In Section\,\ref{sec_i}, we estimate the half-opening angle and the inclination angle of the torus in \src\ by allowing them to vary during spectral fitting. The results suggest an inclination angle smaller than 53$^{\circ}$ and a half-opening angle smaller than 35$^{\circ}$. When we link these two parameters during spectral fitting, the best-fit value corresponds to a small half-opening angle of approximately 26$^{\circ}$.} 

\two{In this scenario, our LOS crosses the upper edge of the torus in \src. Therefore, measured $N_{\rm H, los}$ corresponds to the column density near the edge of the torus, which is only a few per cent of the average column density of the torus: $N_{\rm H, los}=5.2^{+0.7}_{-0.6}\times10^{23}$\,cm$^{-2}$ in 2007 and $2.0^{+0.4}_{-0.3}\times10^{23}$\,cm$^{-2}$ in 2019. A similar  $N_{\rm H, los}$ was found in the \xmm\ observation of \src\ in 2001 \citep[$N_{\rm H, los}\approx3\times10^{23}$\,cm$^{-2}$, ][]{grupe04}. These observations were separated by 18 years and suggest that some variability of  $N_{\rm H, los}$ with a small amplitude may exist in \src\ during this period.}

\two{As shown in the previous section, radiation pressure-driven dusty wind may be forming near the edge of the torus when $N_{\rm H}<10^{24}$\,cm$^{-2}$. If true, our LOS intercepts with the wind and some variability of $N_{\rm H, los}$ is expected. In comparison, the average column density of the torus is as high as $10^{25}$\,cm$^{-2}$. The radiation pressure from the central region of the AGN in \src\ is not high enough to `blow' wind from the equatorial plane of torus.}

\two{Alternatively, a clumpy torus model may also explain the variability in $N_{\rm H, los}$. In the unification paradigm of AGNs \citep{antonucci93}, Seyfert 1 AGNs (Sy1s) and Seyfert 2 AGNs (Sy2s) are distinguished by their inclination angles, i.e. whether our LOS crosses the torus. This standard picture assumes a homogeneous dusty torus. However, observations suggest that the LOS obscuration is determined not only by inclination angle but also by the probability of absorption clouds intercepting our LOS \citep{nenkova08}. Evidence for the clumpiness of the torus includes a large and rapid variability of the absorber observed in several sources \citep[e.g.][]{risaliti02, bianchi12, laha20}. \citet{markowitz14} estimated the probability of an absorption event regardless of constant absorption due to non-clumpy material to be 0.003--0.16 for Sy1s. Dramatic changes of $N_{\rm H, los}$ up to a few orders of magnitude have indeed been seen in Sy1s on observable timescales \citep[e.g.][]{simm18}.}

\two{Unfortunately, archival observations are not sufficient enough to distinguish these two models for the small variability of $N_{\rm H, los}$ in \src. As concluded in the previous section, a search for absorption features, e.g. blueshifted absorption edge due to dusty wind, is required when TES-based observations are available.}

\subsection{\src\ as a Polar-Scattered Sy1}

\subsubsection{The opening angle of the torus in \src}

\two{As introduced above, Sy1s and Sy2s are intrinsically the same type of object but viewed from different inclination angles in the unification paradigm of AGNs \citep{antonucci93}, which is supported by the detection of polarised broad lines in Sy2s \citep[e.g.][]{antonucci85, miller90, young96}. Meanwhile, it is believed that two scattering regions, the equatorial plane \citep[e.g.][]{goodrich94, cohen99, cohen02} and the ionisation cone on the torus axis \citep[e.g.][]{antonucci83,smith02}, both contribute to the polarised emission in AGNs.}

\two{\citet{smith04} argued that the inclination angle plays an important role in the interpretation of the detected polarisation in AGNs. Polar-scattered Sy1s represent a bridge between Sy1s and Sy2s. Their optical polarisation position angle is perpendicular to the projected radio source axis as in Sy2s \citep{smith04}, whilst the majorty of Sy1s show optical polarisation properties that are not consistent with polar scattering \citep[e.g.][]{antonucci83, smith02}. This is because our LOS may intercept with the edge of the torus in polar-scattered Sy1s, e.g. \src\ \citep{smith04}, where the equatorial emission is obscured by the dusty torus and makes little contribution to the detected optical polarisation.}

\two{Our torus modelling of the X-ray spectra of \src\ provides another supporting evidence for this model. Our results suggest that the data are consistent with the case that our LOS intercepts with the upper edge of the torus.} 

\two{Assuming a toroid-shaped inhomogenous torus, we obtain an upper limit of 35$^{\circ}$ for the half-opening angle parameter at 3$\sigma$ uncertainty level (see Section\,\ref{sec_i}). Such a low half-opening angle corresponds to a very high covering factor for \src, e.g. Cfact>0.81.} 

\two{A similar covering factor was found in Mrk~231, another polar-scattered Sy1. For instance, \citet{piconcelli13} found that Mrk~231 shows two partial-covering absorbers with $N_{\rm H}\approx10^{22}$ and $10^{24}$\,cm$^{-2}$. A very high covering factor of more than $0.9$ and $0.8$ are respectively found for these two absorbers \citep{piconcelli13}. Interestingly, \citet{smith04} found that the optical polarisation spectra of \src\ and Mrk~231 are very similar too: their continuum polarisation rises at shorter wavelengths and peaks at the red wing of their broad lines. } 

\subsubsection{Comparison with other polar-scattered Sy1s}

\two{Polar-scattered Sy1s, including \src, often show significant absorption features in the X-ray band, which is in agreement with the picture that our LOS passes close to their torus opening angles \citep[e.g.][]{jimnez08, piconcelli13, laha11, newman21}. For instance, the polar-scattered Sy1 Mrk~704 shows not only a partial covering neutral absorber with $N_{\rm H}\approx10^{23}$\,cm$^{-2}$ and Cfact=0.22 but also two layers of warm absorbers that are associated with the broad line region \citep{laha11}.}

\two{Meanwhile, polar-scattered Sy1s show a variety of X-ray spectral features. For example, the neutral absorber in Mrk~704 has a much lower covering factor than those in \src\ and Mrk~231. This indicates that the torus opening angles in polar-scattered Sy1s may not be uniform. \citet{ricci17b} applied a torus model to a sample of \swift\ BAT AGNs. For the 12 objects with constrained torus half-opening angles, they found a median value of $58\pm3^{\circ}$. \src\ and Mrk~704 are respectively located in the lower and higher ends of the global distribution of torus opening angles.} 

\two{Besides, as demonstrated in this work and \citet{piconcelli13}, Mrk~231 and \src\ show no significant evidence of reprocessing emission from the innermost region, e.g. disc reflection (see Section\,\ref{relxill}). In comparison, other polar-scattered Sy1s show evidence of either broad Fe~K$\alpha$ emission \citep[e.g. Fairall~51,][]{svoboda15} or soft excess emission from the innermost accretion region \citep[e.g. NGC~3227 and Mrk~704,][]{laha11,newman21}.} 

\section{Conclusions}

\two{We present a torus model for the X-ray spectra of the NLS1 \src\ based on archival \suzaku, \nustar\ and \swift\ observations. The main results are as follows.}

\two{
\begin{itemize}
    \item The primary X-ray continuum of \src\ is described by a power law with slope $\Gamma=2.6-3.0$. Such a soft continuum suggests that \src\ is one of the most extreme AGNs that are accreting near or around the Eddington limit. By applying a correction factor to its X-ray luminosity, we obtain $\lambda_{\rm Edd}=0.1-2.8$ after taking into account the uncertainty of the BH mass measurements.
    \item At such a high accretion rate, the radiation pressure from the central region of the AGN may drive wind near the edge of the torus where column density is around $10^{23}$\,cm$^{-2}$. Future high-S/N, TES-based X-ray observations may reveal more spectral details of \src, e.g. blueshifted absorption edge features.
    \item The X-ray data of \src\ are consistent with the optical polarisation model for polar-scattered Sy1s where our LOS intercepts with the upper edge of the torus. The LOS column density of \src\ is only a few per cent of the average column density of the torus in this source. The half-opening angle of the torus is estimated to be around 26$^{\circ}$, corresponding to a very high covering factor of 90\%. Such a small opening angle makes \src\ near the lower end of the global torus opening angle distribution of AGNs.
\end{itemize}
}
\section*{Acknowledgements}

This paper was written during the worldwide COVID-19 pandemic in 2020--2021. We acknowledge the hard work of all the health care workers around the world. We would not be able to finish this paper without their protection. J.J. acknowledges support from the Tsinghua Shui'Mu Scholar Program and the Tsinghua Astrophysics Outstanding Fellowship. M.B. acknowledges support from the YCAA Prize Postdoctoral Fellowship. This work made use of data from the \nustar\ mission, a project led by the California Institute of Technology, managed by the Jet Propulsion Laboratory, and funded by NASA, and data obtained from the Suzaku satellite, a collaborative mission between the space agencies of Japan (JAXA) and the USA (NASA). This research has made use of the \nustar\ Data Analysis Software (NuSTARDAS) jointly developed by the ASI Science Data Center and the California Institute of Technology. 

\section*{Data Availability}

All the data can be downloaded from the HEASARC website at https://heasarc.gsfc.nasa.gov.

%%%%%%%%%%%%%%%%%%%%%%%%%%%%%%%%%%%%%%%%%%%%%%%%%%



\bibliographystyle{mnras}
\bibliography{mrk1239.bib} % if your bibtex file is called example.bib

% \begin{thebibliography}{99}
% \bibitem[\protect\citeauthoryear{Author}{2012}]{Author2012}
% Author A.~N., 2013, Journal of Improbable Astronomy, 1, 1
% \bibitem[\protect\citeauthoryear{Others}{2013}]{Others2013}
% Others S., 2012, Journal of Interesting Stuff, 17, 198
% \end{thebibliography}

%%%%%%%%%%%%%%%%%%%%%%%%%%%%%%%%%%%%%%%%%%%%%%%%%%

%%%%%%%%%%%%%%%%% APPENDICES %%%%%%%%%%%%%%%%%%%%%

\appendix

\section{MYTORUS model} \label{mytorus}

\begin{figure*}
    \centering
    \includegraphics[width=17cm]{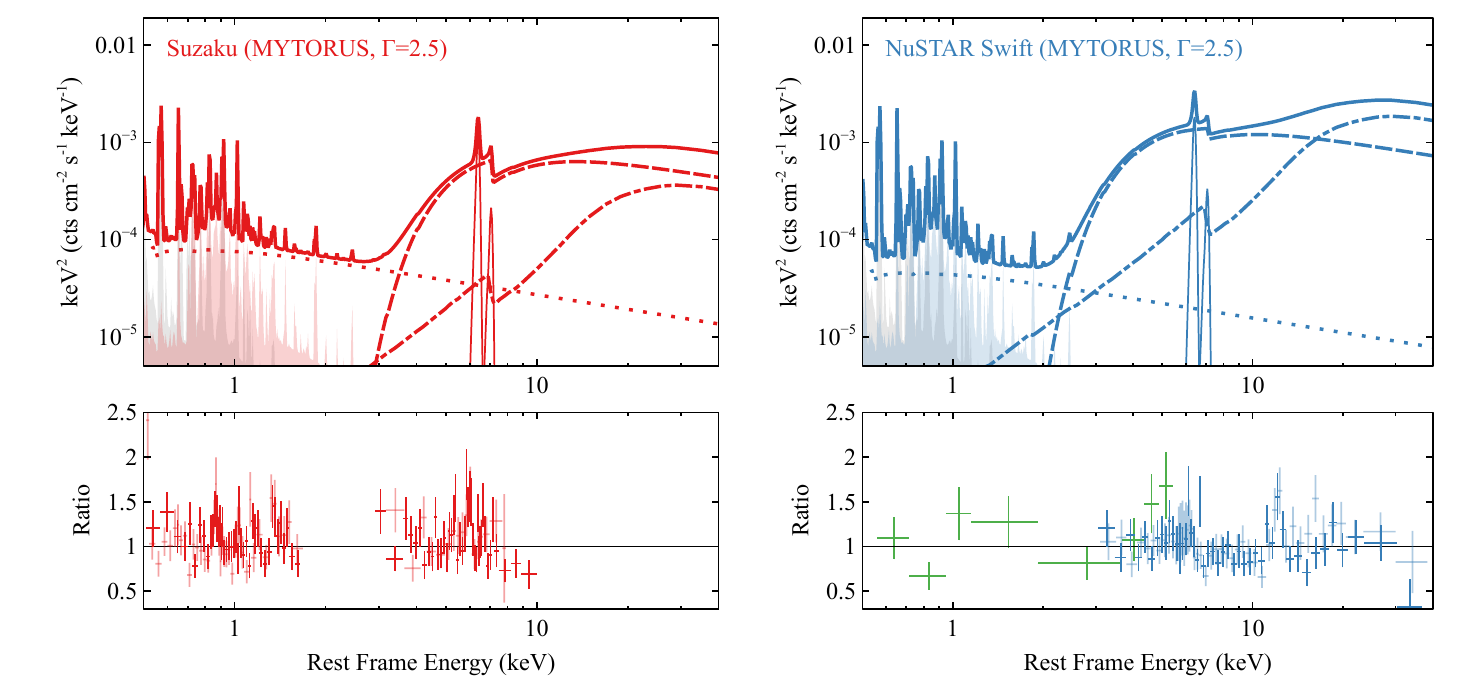}
    \caption{Top: best-fit models for the \suzaku\ (left) and \nustar\ and \swift\ (right) spectra of \src\ using the \texttt{mytorus} model. Thick lines: total model; shaded regions: two diffuse hot plasma emission components; dash-dotted lines: Compton-scattering continuum; dashed lines: absorbed power-law continuum emission; dotted lines: scattered power-law component; thin solid lines: torus line spectra Bottom: corresponding data/model ratio plots. Due to the limited range of $\Gamma$ in the public \texttt{mytorus} model, the photon index is pegged at 2.5 during the analysis of both two sets of data.}
    \label{pic_mytorus}
\end{figure*}

\red{Previously, the \texttt{mytorus} model \citep{murphy09} has been extensively used for spectral modelling of nearby obscured AGNs. In the \texttt{mytorus} model, the covering factor of a toroidal-shaped torus is fixed at 50\% of the sky (a half-opening angle of $60^{\circ}$). This limitation causes inconsistency of the normalisations of the reprocessed spectrum, including the Compton scattering emission and line spectra, with the torus geometry. This decoupling problem is, however, required for high-signal-to-noise broad band data \citep[e.g.][]{guainazzi16}.} 

\red{Another limitation of the available \texttt{mytorus} grids\footnote{The models are available for downloads on the  \texttt{mytorus} website at http://mytorus.com.} is the  range of the photon index parameter. The hard upper limit of the photon index in the public version of \texttt{mytorus} is 2.5, which is too low for the X-ray spectra of \src. }

\red{In this appendix, we present analysis using the \texttt{mytorus} model. The full model is \texttt{constant1 * tbnew * zmshift * ( vmekal1 + vmekal2 +  mtab(trans.abs) * cutoffpl1 + constant2 * cutoffpl2 + constant3 * atab(scattered) + constant4 * atab(fluor) )} in the XSPEC format. The \texttt{tbnew} model accounts for Galactic absorption. The \texttt{zmshift} model accounts for source redshift. The \texttt{vmekal1,2} model calculates the emission from diffuse hot plasma as in MO4. The \texttt{trans.abs} model calculates photoelectric-absorbed zeroth-order continuum. The \texttt{scattered} model calculates Compton-scattering emission of the torus. The \texttt{fluor} accounts for fluorescence emissions for neutral Fe K$\alpha$, Fe K$\beta$ and their Compton shoulders. The normalisations of the \texttt{cutoffpl1,2} models are linked. The coupled mode of the \texttt{mytorus} model is used. The \texttt{constant3} and \texttt{constant4} models, which account for the scaling factors for the scattered emission and the fluorescent line emission respectively, are linked in our analysis.}

\red{The best-fit parameters are shown in Table \ref{tab_mytorus}, and the corresponding models are shown in Fig.\,\ref{pic_mytorus}. The \texttt{mytorus} model provides a much worse fit to the data than MO4 -- our fit shows that the data requires $\Gamma$ to be higher than 2.5. Significant residuals are shown around 6--10\,keV band as shown by the \nustar\ data/model ratio plot. }

\red{Despite the limited range of $\Gamma$, the best-fit \texttt{mytorus} model suggests  $N_{\rm H, ave}=8^{+4}_{-2}\times10^{24}$\,cm$^{-2}$, which is consistent with the value given by the \texttt{borus02} model ($\log(N_{\rm H, ave})=25.0\pm0.2$). The line-of-sight column density $N_{\rm H, los}$ is slightly lower than the values measured by the \texttt{borus02} model due to an improper $\Gamma$.}

\red{The covering factor of the torus in the \texttt{mytorus} model is fixed. The inclination angle of the torus is not constrained by the \texttt{mytorus} model. A 90\% confidence upper limit of $70^{\circ}$ is obtained. In comparison, the upper limit of the inclination angle given by the \texttt{borus02} model is 52$^{\circ}$ when both $i$ and Cfact parameters are free to vary (see Section \ref{sec_i} for more information). } 

\begin{table}
    \centering
    \begin{tabular}{ccccc}
    \hline\hline
    Model & Parameter & Unit & Su & Nu\&Sw \\
    \hline
    \texttt{tbnew} & $N_{\rm H}$ & $10^{20}$\,cm$^{-2}$ & $3.7\pm0.2$ & $l$\\
    \hline
    \texttt{trans.abs} & $N_{\rm H, los}$ & $10^{22}$\,cm$^{-2}$ & $36\pm3$ & $17\pm2$ \\
                        & norm & $10^{-3}$ & $3.3\pm0.3$ & $5.5\pm0.6$ \\
    \hline
    \texttt{scattered} & $N_{\rm H, ave}$ & $10^{24}$\,cm$^{-2}$ & $8^{+4}_{-2}$ & $l$ \\
                    & $E_{\rm cut}$ & keV & 400 & $l$\\
                    & $i$ & deg & $<70$ & $l$ \\
                    & $\Gamma$ & - & >2.49 & >2.49 \\
    \hline
    \texttt{constant3} & $A_{\rm S}$ & - & $17^{+7}_{-10}$ & $15\pm9$ \\
    \hline
    \texttt{constant4} & $A_{\rm L}$ & - & $=A_{\rm S}$ & $=A_{\rm S}$ \\
    \hline
    \texttt{constant2} & $f_{\rm S}$ & $10^{-3}$ & $2.6^{+0.5}_{-0.4}$ & $8.40^{+0.10}_{-0.07}$ \\
    \hline
    \texttt{vmekal2} & kT & keV & $0.58^{+0.06}_{-0.05}$ & $l$ \\
                   & $Z_{\rm Ne}$ & $Z_{\odot}$ & $4.3^{+2.2}_{-1.4}$ & $l$\\
                   & norm & $10^{-5}$ & $4.5^{+1.5}_{-1.2}$ & $l$ \\
    \hline
    \texttt{vmekal1} & kT & keV & $0.182^{+0.017}_{-0.016}$ & $l$ \\
                   & norm & $10^{-4}$ & $1.6\pm0.3$ & $l$ \\
    \hline
    & $\chi^{2}/\nu$ & & 562.69/469 \\
    \hline\hline
    \end{tabular}
    \caption{Best-fit model parameters obtained by fitting the reprocessed emission using \texttt{mytorus}. The half-opening angle of the torus is fixed at $60^{\circ}$ in this model. The high energy cutoff of power-law continuum emission is fixed at 400\,keV. The scaling factor for the fluorescent line emission $A_{\rm l}$ is coupled with that for the scattered continuum $A_{\rm S}$. $f_{\rm S}$ is the scattering fraction due to optically-thin matters outside the LOS.}
    \label{tab_mytorus}
\end{table}

\section{Contribution of disc reflection in the X-ray band} \label{relxill}

\begin{figure}
    \centering
    \includegraphics{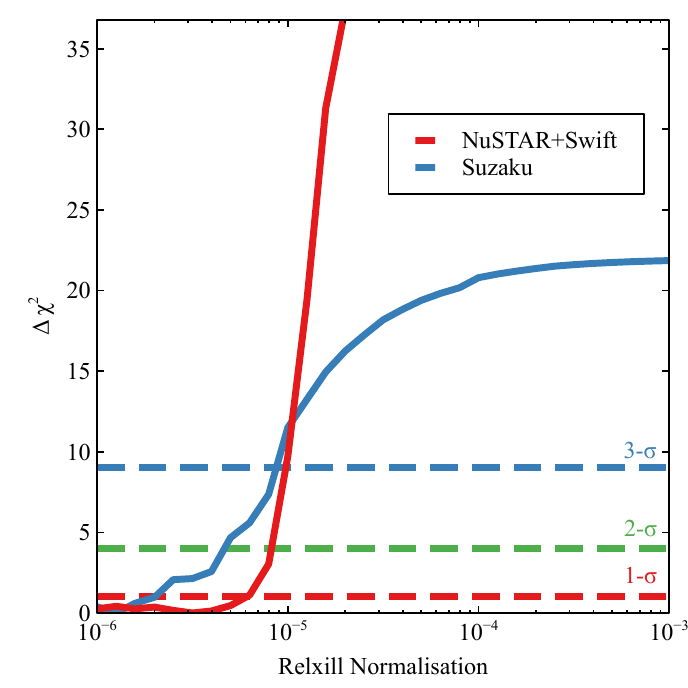}
    \caption{$\chi^{2}$ distributions for the normalisation parameter of the \texttt{relxill} model. Only upper limits are obtained. All the spectra are fitted together. The normalisation parameter is allowed to be different for different epochs. The red solid line shows the measurement using the \nustar\ and \swift\ observations. The blue solid line shows the measurement using the \suzaku\ observation. The dashed lines show 1$\sigma$, 2$\sigma$ and 3$\sigma$ uncertainty ranges.}
    \label{pic_relxill}
\end{figure}

\red{Despite the obscured X-ray emission, the power-law component from the hot corona in the AGN of \src\ is like the one in a typical high-$\lambda_{\rm Edd}$ NLS1 \citep{gallo18}: a very soft power-law index of $\Gamma=2.6-3.0$ is suggested by the data. Reflection off the innermost accretion disc region has been commonly seen in the X-ray observations of other unobscured NLS1s. Prominent features of disc reflection spectra include broad Fe K$\alpha$ emission and Compton hump in the hard X-ray band. In particular, Seyfert 1 AGNs often show excess emission in the soft X-ray band, which could be part of disc reflection as well \citep[e.g.][]{jiang19b}. A strong supporting evidence of soft excess emission being reflection is the increasing number of discoveries of reverberation lags in the soft X-ray band \citep[e.g.][]{kara16}. They are similar to the iron emission and Compton hump reverberation lags \citep[e.g.][]{kara15, kara16}.}

\red{The soft X-ray emission from the centre of the AGN in \src\ is unfortunately obscured by line-of-sight column density. The iron band of \src\ is dominated by fluorescence emissions of the torus. But some disc reflection may still exist \citep{buhariwalla20} and make contribution in the hard X-ray band, the observed Compton hump in particular.} 

\red{In this section, we estimate the upper limit of the contribution from a disc reflection component in the X-ray band. We add the \texttt{relxill} model \citep{garcia13,dauser16} to MO4. The \texttt{relxill} calculates reflection spectra of a relativistic thin disc. We use a power-law emissivity profile for the disc parametrised by an index $q$. The inner disc inclination angle is linked to the inclination angle of the torus, although they might be different in reality. The Ecut and $\Gamma$ parameters of \texttt{relxill} are linked to the corresponding parameters in \texttt{borus02}. Other free parameters include the normalisation parameter, the disc ionisation and the spin of the BH. The inner radius of the disc is assumed to be at the innermost stable circular orbit. By doing so, we only obtain an upper limit of the \texttt{relxill} component for both the \suzaku\ and the \nustar\ epochs. See Fig.\,\ref{pic_relxill} for the constraints of the normalisation parameters of \texttt{relxill} for two epochs.}

\red{The 3$\sigma$ upper limit of the normalisation parameters of \texttt{relxill} is approximately $10^{-5}$ for both epochs. At the 3$\sigma$ upper limit, the \texttt{relxill} component produces an observed flux of approximately $3\times10^{-14}$\,\ergps\ in the 3--10\,keV band for the \suzaku\ epoch and an observed flux of around $6.8\times10^{-13}$\,\ergps\ in the 3--78\,keV band for the \nustar\ epoch. They respectively take up  4\% and 8\% of the total observed X-ray flux in corresponding total energy bands\footnote{The \texttt{relxill} component takes up to 6\% and 9\% in the 10--40\,keV band respectively for the \nustar\ and \suzaku\ epochs.}. We conclude that the narrow Fe K$\alpha$ emission and the Compton hump shown in the X-ray observations are likely to be dominated by reprocessed emission from the torus based on the obscured X-ray nature and observations in other wavelengths \citep{smith04, rodriguez06}. The inner disc reflection makes little contribution to the data with a 3$\sigma$ upper limit of 8\% during the \nustar\ epoch.}

\red{The lack of evidence for broad Fe K$\alpha$ emission is interesting. Following are possible explanations: 1) the disc might be very ionised. As shown by \citet{jiang20}, the discs in some very extreme, unobscured NLS1s with $\Gamma>2.5$ show a high ionisation state of $\log(\xi)>3$. In comparison, typical Seyfert AGNs have a lower-ionisation disc of $\log(\xi)\approx1-2$ \citep{walton13}. At such a high ionisation state, the Fe K$\alpha$ emission is weak \citep[e.g.][]{garcia13}. 2) The high-$\lambda_{\rm Edd}$ NLS1s in \citet{jiang20} and \src\ are all very soft X-ray emitters. The S/N of the current X-ray CCD data, e.g. from \xmm\ or \suzaku, is not high enough to detect their broad Fe K$\alpha$ emission. Future high-S/N observations, e.g. from \textit{Athena}, might be able to do so according to the simulations in \citet{jiang20}. 3) The innermost accretion region may not hold a thin disc geometry at a high Eddington ratio as in \src. Due to the thickness of an optically thick disc, emissions from the innermost region might be obscured by the puffed-up disc \citep[e.g.][]{ken05}. Consequently, a lower reflection fraction is expected for an optically thick, geometrically slim disc with a compact coronal region in comparison with a geometrically thin disc.}
 
% \section{Srepome extra material}

% If you want to present additional material which would interrupt the flow of the main paper,
% it can be placed in an Appendix which appears after the list of references.

%%%%%%%%%%%%%%%%%%%%%%%%%%%%%%%%%%%%%%%%%%%%%%%%%%

% Don't change these lines
\bsp	% typesetting comment
\label{lastpage}
\end{document}